\documentclass[preprint,showpacs,floatfix]{revtex4}
\usepackage[dvips]{graphicx}
\begin{document}
\title{$N_pN_n$ dependence of empirical formula for the lowest \\excitation energy of the $2^+$ states in even-even nuclei}
\author{Guanghao \surname{Jin}}
\author{Jin-Hee \surname{Yoon}}
\author{Dongwoo \surname{Cha}}
\email{dcha@inha.ac.kr}
\thanks{Fax: +82-32-866-2452}
\affiliation{Department of Physics, Inha University, Incheon
402-751, Korea}
\date{December 12, 2007}

\begin{abstract}
We examine the effects of the additional term of the type $\sim e^{- \lambda' N_pN_n}$ on the recently proposed empirical formula for the lowest excitation energy of the $2^+$ states in even-even nuclei. This study is motivated by the fact that this term carries the favorable dependence of the valence nucleon numbers dictated by the $N_pN_n$ scheme. We show explicitly that there is not any improvement in reproducing $E_x(2_1^+)$ by including the extra $N_pN_n$ term. However, our study also reveals that the excitation energies $E_x(2_1^+)$, when calculated by the $N_pN_n$ term alone (with the mass number $A$ dependent term), are quite comparable to those calculated by the original empirical formula.
\end{abstract}


\maketitle

Recently, the role played by the valence nucleon numbers $N_p$ and $N_n$ in determining the lowest excitation energy of the natural parity even multipole states in even-even nuclei has been emphasized \cite{Kim}. Over the past four decades, the valence nucleon numbers $N_p$ and $N_n$ have been frequently adopted in phenomenologically parameterizing various nuclear observables. It was Hamamoto who first pointed out that the square roots of the ratios of the measured and the single particle $B(E2)$ values were proportional to the product $N_pN_n$ \cite{Hamamoto}. Subsequently, it has been shown that a very simple pattern emerged when the nuclear data on the lowest excited states in the nuclei was plotted in terms of the product $N_pN_n$ \cite{Casten1}. For example, the measured excitation energies $E_x(2_1^+)$ of the lowest excited $2^+$ states in even-even nuclei make up the scattered irregular points when they are plotted against the mass number $A$ ($A$-plot). However, they become very neatly rearranged points when they are plotted against the product $N_pN_n$ ($N_pN_n$-plot) \cite{Yoon}. A similar simplification was observed from the ratio $E_x(4_1^+)/E_x(2_1^+)$ \cite{Casten3,Casten4,Cakirli}, the transition probability
$B(E2; 2_1^+ \rightarrow 0^+)$ \cite{Casten5}, and the quadrupole
deformation parameter $e_2$ \cite{Zhao}. This interesting phenomenon has been called the $N_pN_n$ scheme in the literature. In fact, the $N_pN_n$ scheme has been extensively and successfully used for more than two decades to correlate large amounts of data on the collective degrees of freedom in nuclei \cite{Casten2}. It has also been established that the $N_pN_n$ scheme is manifested because the valence proton-neutron (p-n) interaction is the dominant controlling factor in the development of collectivity in nuclei \cite{Heyde,Federmann,Dobaczewski}. In fact, most recently the measured p-n interaction distribution in even-even nuclei over the entire nuclear chart has been successfully accounted for by the nuclear density functional theory calculations \cite{Stoitsov}.

In a previous publication \cite{Ha1}, we proposed an empirical formula in the form of
\begin{equation} \label{E}
E_x = \alpha A^{-\gamma} + \beta \left[ e^{- \lambda N_p} +
e^{- \lambda N_n} \right],
\end{equation}
for the excitation energy $E_x(2_1^+)$ of the first $2^+$ states in even-even nuclei. The model parameters $\alpha$, $\beta$, $\gamma$, and $\lambda$ can be fixed easily and unambiguously by the usual least squares fitting technique. However, since the excitation energies span a broad range and sometimes the difference between the measured and calculated excitation energies can be too large, we take the logarithmic error $R_E(i)$, for the $i$th data point, of the calculated excitation energy $E_x^{\rm cal} (i)$ with respect to the measured excitation energy $E_x^{\rm exp} (i)$ by \cite{Sabbey}
\begin{equation} \label{error}
R_E(i) = \log \left[ { E_x^{\rm cal}(i) \over E_x^{\rm exp}(i)} \right] = \log \left[ E_x^{\rm cal}(i) \right] - \log \left[ E_x^{\rm exp}(i) \right].
\end{equation}
Then we minimize the dimensionless $\chi^2$ value which is defined in terms of the logarithmic error by
\begin{equation} \label{Chi}
\chi^2 = { 1 \over {N_0}} \sum_{i=1}^{N_0} \Big| R_E(i) \Big|^2
\end{equation}
where $N_0$ is the number of total data points considered. The empirical formula, Eq.\,(\ref{E}), has been found to be capable of describing the essential trends of $E_x(2_1^+)$ in even-even nuclei throughout the entire periodic table \cite{Ha1}. We have also shown that the source, which governs the $2_1^+$ excitation energy dependence given by Eq.\,(\ref{E}) on the valence nucleon numbers, is the effective particle number that participates in the residual interaction from the Fermi level \cite{Ha2}. In addition, it is shown that Eq.\,(\ref{E}) complies with the requirement of the $N_pN_n$ scheme, even though it does not depend explicitly on the product $N_pN_n$ \cite{Yoon}. Furthermore, the same formula could be successfully adopted in describing the lowest excitation energy of the natural parity even multipole states up to $10^+$ in even-even nuclei \cite{Kim}.

In spite of the fact that the above empirical formula complies well with the $N_pN_n$ scheme, there still remains some possibility that the excitation energy could depend explicitly on the product $N_pN_n$ as dictated by the $N_pN_n$ scheme, which assumes the active contribution of the valence p-n interaction. In this work, therefore, we want to include an additional term of the type $\sim e^{- \lambda' N_pN_n}$ ($N_pN_n$ term) in Eq.\,(\ref{E}) and to check how this term competes with other terms in reproducing the lowest $2^+$ excitation energy $E_x(2_1^+)$ in even-even nuclei. The most general form of the proposed formula with the additional $N_pN_n$ term can be written as:
\begin{equation} \label{E1}
E_x = \alpha A^{-\gamma} + \beta_p e^{- \lambda_p N_p} + \beta_n e^{- \lambda_n N_n} + \beta' e^{- \lambda' N_pN_n}
\end{equation}
where the parameters $\beta$ and $\lambda$ in Eq.\,(\ref{E}) are split into $\beta_p$, $\beta_n$, and $\lambda_p$, $\lambda_n$, respectively. This takes into account the possibility that the contributions to the excitation energy $E_x$ from protons and neutrons might be different. The parameters in Eq.\,(\ref{E1}) are determined by minimizing the $\chi^2$ value given by Eq.\,(\ref{Chi}). Because the fitting procedure adopting the $\chi^2$ value in terms of the logarithmic error is quite competent, we anticipate naively that, if the $N_pN_n$ term contributes more effectively than the separate $N_p$ and $N_n$ terms in Eq.\,(\ref{E1}), then the fitting procedure would return relatively smaller values for the parameters $\beta_p$ and $\beta_n$ than for $\beta'$ in Eq.\,(\ref{E1}).

We perform the fitting procedure under the following five different cases, according to the way in which the parameters in Eq.\,(\ref{E1}) are treated: (I) $\beta_p=\beta_n=\beta$ and $\lambda_p=\lambda_n=\lambda$ with $\beta'=0$ and $\lambda'=0$ (4+0 parameters); (II) $\beta_p=\beta_n= \beta$ and $\lambda_p=\lambda_n=\lambda$ (4+2 parameters); (III) $\beta'=0$ and $\lambda'=0$ (6+0 parameters); (IV) no restriction (6+2 parameters) and; (V) $\beta_p=\beta_n=0$ and $\lambda_p=\lambda_n=0$ (2+2 parameters). We have distinguished the above cases by labeling them with two digits connected by a plus sign such as `$n_1 + n_2 $ parameters', as shown in the above parenthesis. The first digit denotes the number of parameters adopted from the first three terms in Eq.\,(\ref{E1}), while the second digit denotes the number of parameters adopted from the last term of the same equation.

\begin{table}[b]
\begin{center}
\caption{Values for the parameters in Eq.\,(\ref{E1}) for the excitation energy of the first $2^+$ state determined by minimizing $\chi^2$ value under the five cases defined in the text.}
\begin{tabular} {lrrrrrrr}
\hline\hline
Case&I&II&II$'$&III&IV&IV$'$&V\\
Parameters~~~~~&~~~~~4+0&~~~~~4+$2_1$&~~~~~4+$2_2$&~~~~~6+0&~~~~~6+$2_1$ &~~~~~6+$2_2$&~~~~~2+2\\
\hline
$\alpha$(MeV)&81.39&69.65&11.00&68.38&64.31&14.45&68.82\\
$\gamma$&1.375&1.331&0.685&1.342&1.318&0.795&1.319\\
$\beta_p$(MeV)&0.958&0.391&0.922&0.826&0.399&0.796&$\,$\\
$\beta_n$(MeV)&0.958&0.391&0.922&1.167&0.754&1.141&$\,$\\
$\lambda_p$&0.344&$\infty$&0.310&0.419&0.750&0.359&$\,$\\
$\lambda_n$&0.344&$\infty$&0.310&0.284&0.374&0.284&$\,$\\
$\beta'$(MeV)&$\,$&0.795&-0.311&$\,$&0.510&-0.216&0.964\\
$\lambda' \times 10^4$&$\,$&233&9&$\,$&215&11&271\\
\hline
$\chi^2$&0.151&0.128&0.126&0.126&0.117&0.109&0.140\\
\hline\hline
\end{tabular}
\label{tab-1}
\end{center}
\end{table}

The resulting parameter values for each case are listed in Tab.\,\ref{tab-1}, together with the corresponding dimensionless $\chi^2$ values. Note, however, that the definition of the $\chi^2$ value given by Eq.\,(\ref{Chi}) is unrelated to one usually employed in the standard error analysis. Therefore the $\chi^2$ values shown in Tab.\,\ref{tab-1} should be referred for the purpose of comparison only. The measured excitation energies $E_x(2_1^+)$ in this calculation are quoted from the compilation in Raman {\it et al}. \cite{Raman}. When the $N_pN_n$ term is not included in Eq.\,(\ref{E1}), a unique parameter set exists, which minimizes the $\chi^2$ value. The cases of I (4+0 parameters) and III (6+0 parameters) in Tab.\,\ref{tab-1}, both correspond to this type of situation. But, we have noticed that when the $N_pN_n$ term is present in Eq.\,(\ref{E1}), there are two parameter sets, both of which constitute two local minima for the $\chi^2$ value. These two parameter sets are distinguished in Tab.\,\ref{tab-1} by adding a subscript to the second digit (e.g., 4+$2_1$ and 4+$2_2$).  The cases of II (4+$2_1$ and 4+$2_2$ parameters) and IV (6+$2_1$ and 6+$2_2$ parameters) correspond to this type of situation. However, when only the $N_pN_n$ term is present in Eq.\,(\ref{E1}), such as in case V in Tab.\,\ref{tab-1}, there is a recurrence of a unique parameter set.

\begin{figure}[b]
\centering
\includegraphics[width=11.0cm,angle=0]{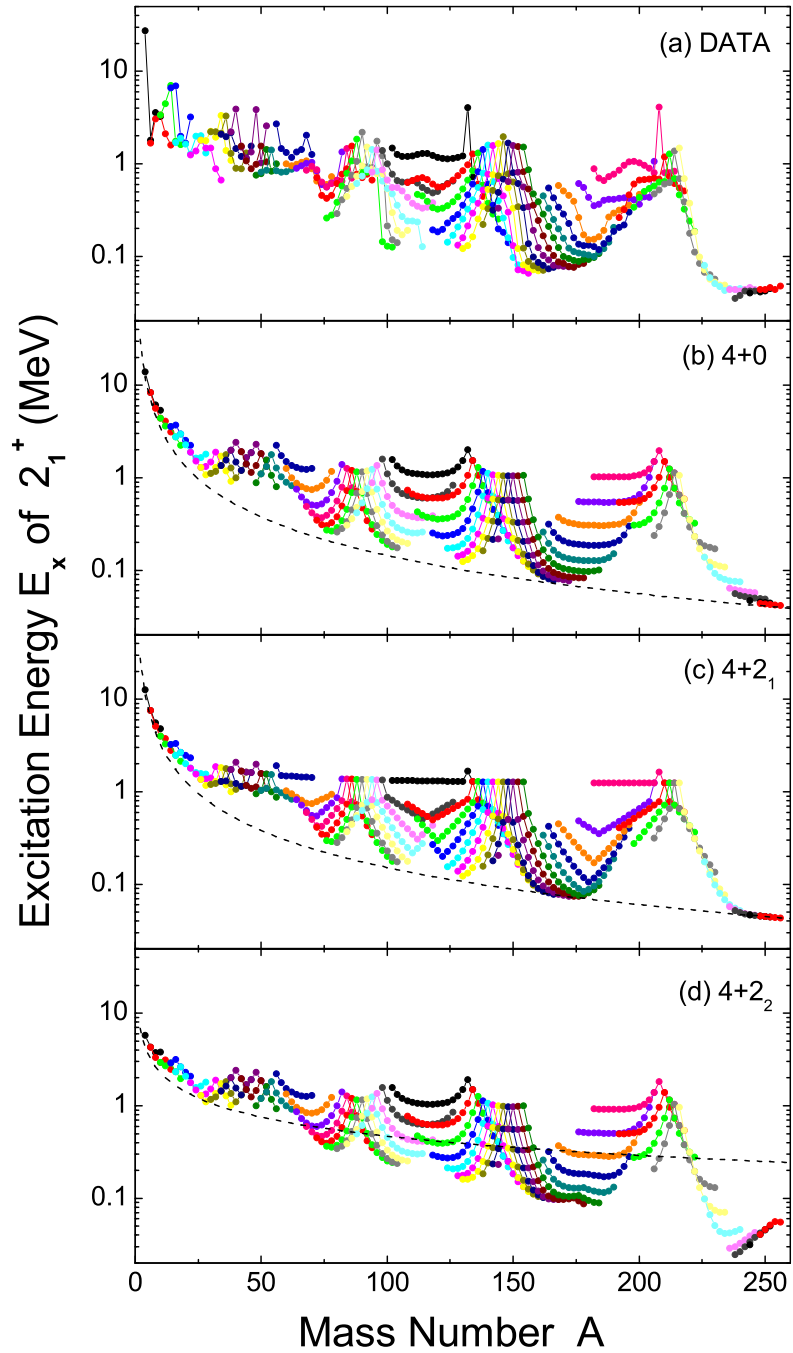}
\caption{Excitation energies of the first $2^+$ states in even-even nuclei. The points are connected by solid lines along the isotopic chains. Part (a) shows the measured excitation energies \cite{Raman} while parts (b), (c), and (d) show those calculated by our empirical formula given by Eq.\,(\ref{E1}) with the parameter set given by 4+0, 4+$2_1$, and 4+$2_2$ parameters in Tab.\,\ref{tab-1}, respectively. The dashed lines in parts (b), (c), and (d) are drawn by taking only the first term $\alpha A^{-\gamma}$ of Eq.\,(\ref{E1}).}
\label{fig-1}
\end{figure}

\begin{figure}[b]
\centering
\includegraphics[width=11.0cm,angle=0]{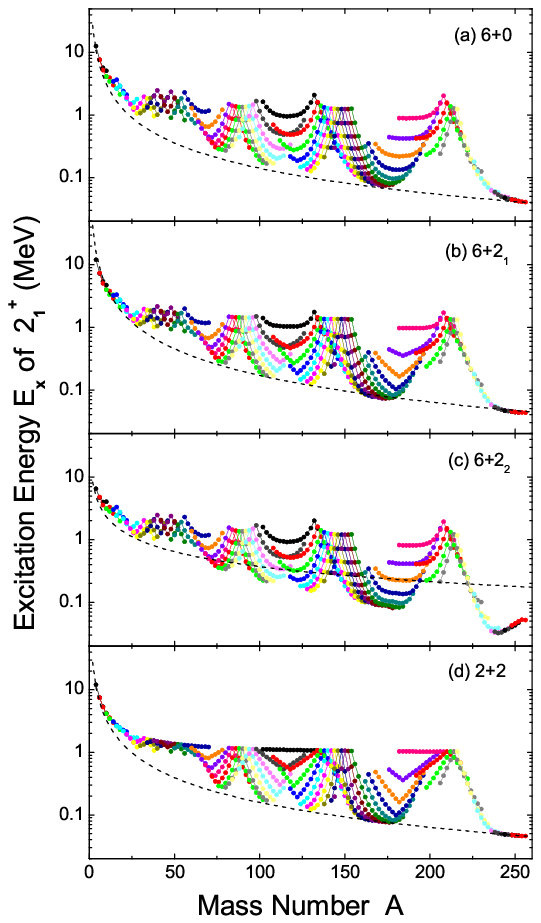}
\caption{Same as Fig.\,\ref{fig-1} but for parts (a), (b), (c), and (d) show the excitation energies calculated by our empirical formula, Eq.\,(\ref{E1}) with the parameter set given by 6+0, 6+$2_1$, 6+$2_2$, and 2+2 parameters in Tab.\,\ref{tab-1}, respectively.}
\label{fig-2}
\end{figure}

Fig.\,\ref{fig-1} presents the results for the first $2^+$ excitation energies $E_x (2_1^+)$ in even-even nuclei obtained by the four-parameter empirical formula, Eq.\,(\ref{E}). The results obtained by the six-parameter empirical formula, Eq.\,(\ref{E1}), are presented in Fig.\,\ref{fig-2}. In both figures, the points connected by solid lines represent nuclei that belong to the same atomic number (the isotopic chains). The dashed lines of the two figures are drawn by taking only the first term $\alpha A^{-\gamma}$ of the empirical formula. The calculated excitation energies are to be compared with the measured data shown in Fig.\,\ref{fig-1}(a).

In Figs.\,\ref{fig-1}(b), 1(c), and 1(d), the $A$-plots of the excitation energies $E_x (2_1^+)$ are shown, which are calculated by adopting the parameter set of the 4+0 parameter for Fig.\,1(b), the 4+$2_1$ parameter for Fig.\,1(c), and the 4+$2_2$ parameter for Fig.\,1(d) as given in Tab.\,\ref{tab-1}. As has been discussed in the previous publications \cite{Kim,Ha1}, we find from Fig.\,1(b) that the four-parameter empirical formula, Eq.\,(\ref{E}), describes reasonably well the essential trends of the lowest $2^+$ excitation energies in even-even nuclei throughout the entire periodic table. We also find, by comparing Figs.\,1(c) and 1(d) with 1(b), that the $N_pN_n$ term added to the four-parameter empirical formula, does not have a significant influence on the overall shape of the graphs. However, we also notice that there are a few small variations brought about by the $N_pN_n$ term. First of all, we find that the $N_pN_n$ term can contribute either additively (as in Fig.\,1(c)) or subtractively (as in Fig.\,1(d)) to the original formula. This fact can be reconfirmed by the opposite sign of the parameter $\beta'$ for the cases II and II$'$ in Tab.\,\ref{tab-1}. It is of notable interest that the dashed curve in Fig.\,1(c) (where the $N_pN_n$ term adds to the remaining terms), follows the bottom contour line of the excitation energies, while the dashed curve in Fig.\,1(d) (where the $N_pN_n$ term subtracts from the remaining terms), passes through the middle of the excitation energies. Even though the $\chi^2$ value for case II$'$ (4+$2_2$ parameters) is lower than it is for case II (4+$2_1$ parameters) by only 0.002, we find that the results obtained by case II reproduce the measured excitation energies better than the results obtained by case II$'$ in the mid-shell region. However, it is interesting to find that, in case II, there is no contribution from the second and third terms in Eq.\,(\ref{E1}), except for $N_p=0$ or $N_n=0$, as a result of the fitted parameter values $\lambda_p=\lambda_n=\infty$. Consequently, an almost straight line represents the $A$-plots of the excitation energies $E_x(2_1^+)$ for $N_p=0$, while the excitation energies $E_x(2_1^+)$ for $N_p=N_n=0$ nuclei are represented by dots above the previous straight line in Fig.\,1(c), which does not concur with the shape given by the data. This eccentric behavior occurs because, under the condition of $\beta_p=\beta_n$ and $\lambda_p=\lambda_N$, the last $N_pN_n$ term can play the role of both the second and third terms in Eq.\,(\ref{E1}).

By comparing case I with case II in Tab.\,\ref{tab-1}, we learn that, when the $N_pN_n$ term is added to the four-parameter empirical formula, the $\chi^2$ value is lowered by approximately $15 \%$, from 0.151 to 0.128. Almost the same amount of reduction in the $\chi^2$ value can be obtained if the four-parameter formula is extended to the six-parameter one. This can be seen from the $\chi^2$ values of  case I and case III in Tab.\,\ref{tab-1}. However, by adding the $N_pN_n$ term to the six-parameter empirical formula (compare case III with case IV), a further reduction in $\chi^2$ value can be achieved, which amounts to only approximately $6 \%$. In Figs.\,2(a), 2(b), and 2(c), we show the $A$-plots of the excitation energies $E_x(2_1^+)$. These are calculated by adopting the parameter set of the 6+0 parameters for Fig.\,2(a), the 6+$2_1$ parameters for Fig.\,2(b), and the 6+$2_2$ parameters for Fig.\,2(c), as given in Tab.\,\ref{tab-1}. The three graphs calculated by the six-parameter empirical formula in Fig.\,\ref{fig-2} are basically same as the corresponding three graphs calculated by the four-parameter counterparts in Fig.\,\ref{fig-1} except for the graph shown in Fig.\,2(b). For case IV, where the four parameters $\beta_p$, $\beta_n$, $\lambda_p$, and $\lambda_n$ are separately fitted, there are suitable contributions from the second and third terms in Eq.\,(\ref{E1}) and the eccentric behavior shown in Fig.\,1(c) disappears, as shown in Fig.\,2(b).

\begin{figure}[b]
\centering
\includegraphics[width=11.0cm,angle=0]{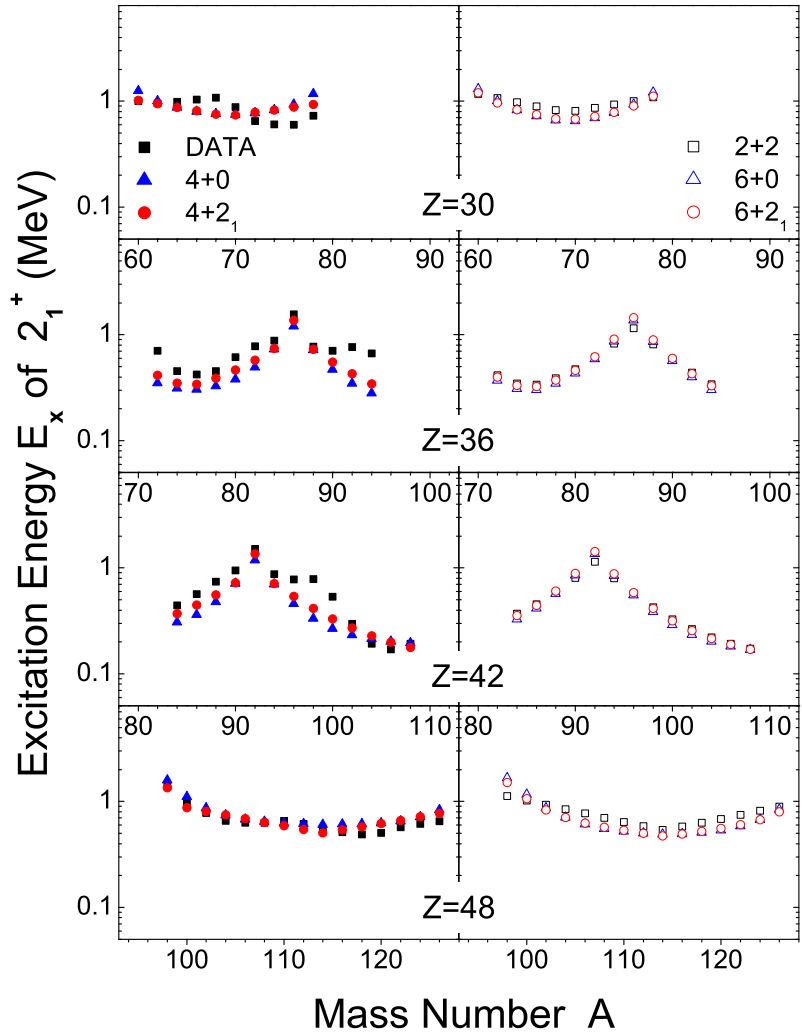}%
\caption{Excitation energies of the first $2^+$ states in nuclei which belong to the isotopic chains $Z=30$, $Z=36$, $Z=42$, and $Z=48$. The measured excitation energies (solid squares) as well as the calculated ones by adopting the 4+0 (solid triangles), 4+$2_1$ (solid circles), 2+2 (empty squares), 6+0 (empty triangles), and 6+$2_1$ (empty circles) parameter sets are shown.} \label{fig-3}
\end{figure}

By reviewing Figs. \ref{fig-1} and \ref{fig-2}, it becomes apparent that the role played by the supplementary $N_pN_n$ term is insignificant when it is added to the four or six parameter empirical formula. However, we present a very interesting results in Fig.\,\ref{fig-2}(d) where we show the $A$-plot of the excitation energies $E_x(2_1^+)$ calculated by only the $N_pN_n$ term together with the first $\alpha A^{-\gamma}$ term in Eq.\,(\ref{E1}). In this figure, we adopt the 2+2 parameter set which is listed as case V in Tab.\,\ref{tab-1}. It is quite surprising to find that the graph obtained by the 2+2 parameter set is highly comparable to that obtained by the 6+$2_1$ parameter set, with the exception of the excitation energies in nuclei that belong to the closed proton and/or the neutron major shell. Furthermore, since the $\chi^2$ value should decrease monotonically as the number of parameters increases, the results of the 2+2 parameter set, where $\chi^2 = 0.140$, could be regarded better than, or at least as good as, the results of the 4+0 parameter set, where $\chi^2 = 0.151$.

\begin{figure}[b]
\centering
\includegraphics[width=11.0cm,angle=0]{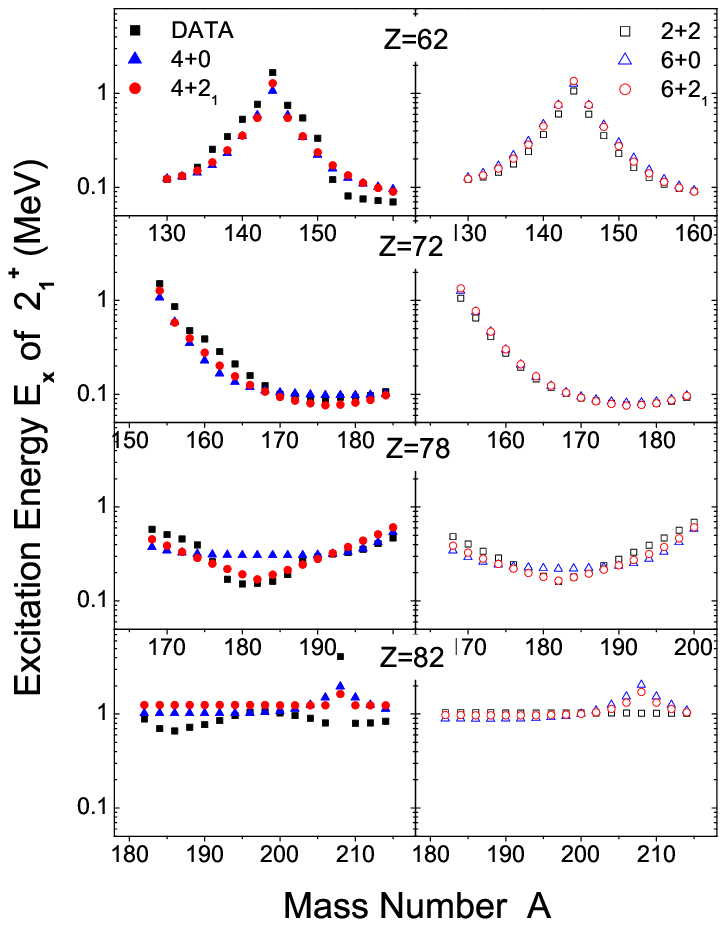}%
\caption{Same as Fig.\,\ref{fig-3} but for the isotopic chains $Z=62$, $Z=72$, $Z=78$, and $Z=82$.} \label{fig-4}
\end{figure}

\begin{figure}[b]
\centering
\includegraphics[width=11.0cm,angle=0]{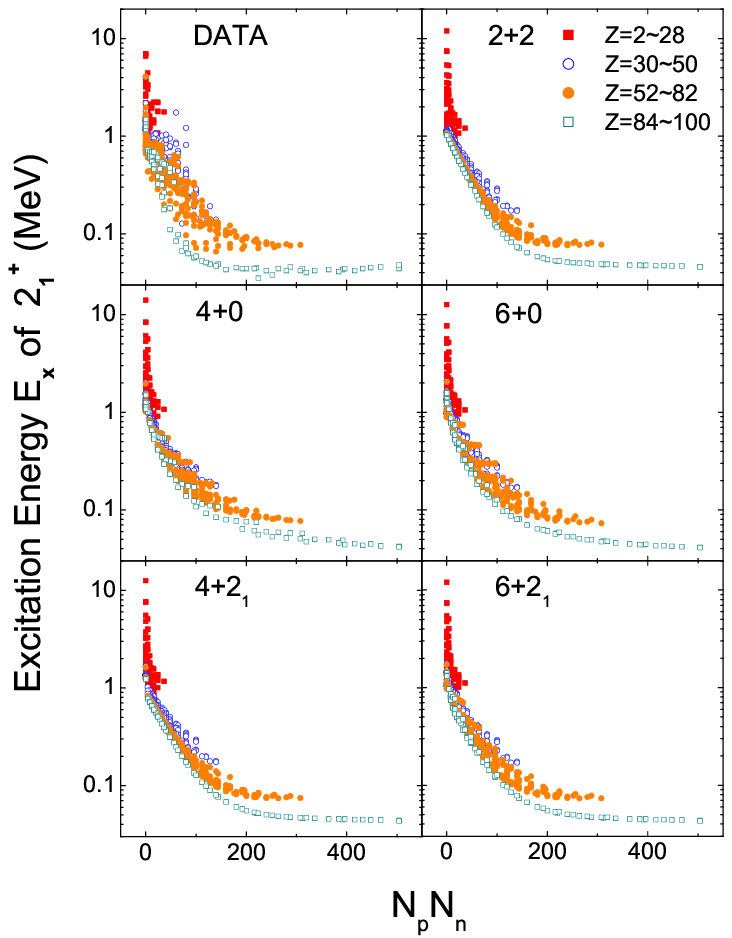}%
\caption{The $N_pN_n$-plot versions of the $A$-plots shown in Fig.\,\ref{fig-1} and \ref{fig-2}. The excitation energies are expressed by the following different symbols according to which major shells they belong to: solid squares ($Z=2 \sim 28$), empty circles ($Z=30 \sim 50$), solid circles ($Z=52 \sim 82$), and empty squares ($Z=84 \sim 100$).} \label{fig-5}
\end{figure}

In order to provide a closer confirmation of the relative performance of the various cases presented in Figs.\,\ref{fig-1} and \ref{fig-2}, we choose a number of individual isotopic chains and remake the $A$-plots of them in Figs.\,\ref{fig-3} and \ref{fig-4}. In Fig.\,\ref{fig-3}, we display arbitrarily chosen four isotopic chains, $Z=30$, $Z=36$, $Z=42$, and $Z=48$, which include nuclei that belong to the proton major shell between $Z=30$ and $50$. The left panels of Fig.\,\ref{fig-3} show the measured excitation energies $E_x(2_1^+)$ (solid squares) as well as those calculated by adopting the 4+0 (solid triangles) and 4+$2_1$ (solid circles) parameter sets. In addition, the right panels of the same figure show the excitation energies $E_x(2_1^+)$, calculated by adopting the 2+2 (empty squares), 6+0 (empty triangles), and 6+$2_1$ (empty circles) parameter sets. These graphs confirm that the addition of the $N_pN_n$ term has a minimal effect in both empirical formulae, Eq.\,(\ref{E}) and Eq.\,(\ref{E1}). Furthermore, it is interesting to discover that the excitation energies $E_x(2_1^+)$ calculated by only the $N_pN_n$ term with the mass number dependent $\alpha A^{-\gamma}$ term (i.e., by the 2+2 parameter set), are as satisfactory as those calculated by any other parameter sets. Similarly in Fig.\,\ref{fig-4}, we display again arbitrarily chosen four isotopic chains, such as $Z=62$, $Z=72$, $Z=78$, and $Z=82$, which include nuclei that belong to the proton major shell between $Z=52$ and $82$. This figure also shows that the excitation energies $E_x(2_1^+)$, calculated by the parameter sets with the additional $N_pN_n$ term, are found to be more or less the same as those that were calculated by the parameter sets without the $N_pN_n$ term. While the differences between these graphs are small, the case with the best results is the 6+0 parameter set. Fig.\,\ref{fig-4} also shows that the results obtained by the 2+2 parameter set are very similar to those obtained by any other parameter sets, with the exception of $Z=82$, which is the proton closed shell.

Finally, we plot the excitation energies $E_x(2_1^+)$ shown in Figs.\,\ref{fig-1} and \ref{fig-2} again in Fig.\,\ref{fig-5} but this time we make the $N_pN_n$-plots instead of the $A$-plots. Of course, the graphs in Fig.\,\ref{fig-5} are drawn with exactly the same set of plotted points as those that were used in Figs.\,\ref{fig-1} and \ref{fig-2}. The three left panels of Fig.\,\ref{fig-5} show the $N_pN_n$-plot versions of the $A$-plots that were shown in Fig.\,\ref{fig-1}(a) (data), \ref{fig-1}(b) (4+0), and \ref{fig-1}(c) (4+$2_1$), while the three right panels show the $N_pN_n$-plot versions of the $A$-plots shown in Figs.\,\ref{fig-2}(d) (2+2), \ref{fig-2}(a) (6+0), and \ref{fig-2}(b) (6+$2_1$). In these $N_pN_n$-plots, the excitation energies $E_x(2_1^+)$ are expressed by the following different symbols, according to which major shell they belong to: solid squares ($Z=2 \sim 28$); empty circles ($Z=30 \sim 50$); solid circles ($Z=52 \sim 82$); and empty squares ($Z=84 \sim 100$). We can observe from these $N_pN_n$-plots (as with the $A$-plots), that the overall characteristics of the data are reasonably reproduced by our empirical formula. We also find from the $N_pN_n$ plots that the $N_pN_n$ term added to the empirical formula does not alter the essential shape of the graph. However, we can still recognize the effect brought by the addition of the $N_pN_n$ term to the empirical formula by comparing the lowest two left panels or the lowest two right panels. The width of the plotted points in the $N_pN_n$-plot becomes narrow and the plotted points which belong to the different major shells become further separated when the $N_pN_n$ term is present, than when the $N_pN_n$ term is absent. However, by comparing calculated results with the data shown in the top left panel of Fig.\,\ref{fig-5}, it is evident that both of these effects, caused by the additional $N_pN_n$ term, deteriorate the fit to the data. Just as pointed out regarding Fig.\,\ref{fig-2}, the $N_pN_n$ plot drawn by the 2+2 parameter set in Fig.\,\ref{fig-5} is very similar to the $N_pN_n$-plot drawn by any other parameter sets. Furthermore, it is identical to the $N_pN_n$-plot drawn by the 4+$2_1$ parameter set. This is because, as discussed previously, the second and third terms of Eq.\,(\ref{E1}) do not contribute in any way to $E_x(2_1^+)$, when this parameter set is adopted.

In summary, we wanted to double check any possibility that the first $2^+$ excitation energy could depend explicitly on the product $N_pN_n$ as expected by the $N_pN_n$ scheme. We, therefore, included the $N_pN_n$ term of the type $\sim e^{- \lambda' N_pN_n}$ as in Eq.\,(\ref{E1}) and investigated how this term competed with other terms in reproducing the first $2^+$ excitation energy $E_x(2_1^+)$ in even-even nuclei. In fact, we anticipated naively that if the $N_pN_n$ term contributed more effectively than the separate $N_p$ and $N_n$ terms in Eq.\,(\ref{E1}), then our fitting procedure would return much smaller values for the parameters $\beta_p$ and $\beta_n$ than $\beta'$ in Eq.\,(\ref{E1}). But contrary to our anticipation, our results show that the role played by the $N_pN_n$ term is insignificant when it is added to the former empirical formula as evidenced by the $A$-plots displayed in Figs.\,\ref{fig-1} and \ref{fig-2}. However, we also end up with an unexpected but very interesting result. The $A$-plot drawn by the 2+2 parameter set (calculated with only the $N_pN_n$ term and the $\alpha A^{-\gamma}$ term) is quite similar to those drawn by any other parameter set. But the result calculated by the 2+2 parameter set has the following two observational flaws. One is that it could not reproduce the excitation energies $E_x(2_1^+)$ in nuclei that belong to the closed proton and/or neutron major shell where $N_p=0$ and/or $N_n=0$. The other is that the $N_pN_n$-plot of the 2+2 parameter set is too narrow in order to match the $N_pN_n$-plot of the data. However, one must not forget the fact that the $\chi^2$ value for the 2+2 parameter set is quite comparable to that for the 4+0 parameter set. In fact, it is fair to say that the present results alone could not tell which one, the $N_p$ cross $N_n$ term or the $N_p$ plus $N_n$ term, would finally represent the lowest excitation energy systematics in nuclei. A further study to clear up this situation is called for.

\begin{acknowledgments}
This work was supported by an Inha University research grant.
\end{acknowledgments}


\begin{references}
\bibitem{Kim} D. Kim, E. Ha, and D. Cha, Nucl. Phys. {\bf A799}, 46 (2008).
\bibitem{Hamamoto} I. Hamamoto, Nucl. Phys. {\bf 73}, 225 (1965).
\bibitem{Casten1} R. F. Casten, Nucl. Phys. {\bf A443}, 1 (1985).
\bibitem{Yoon} J-H Yoon, E. Ha, and D. Cha, J. Phys. G: Nucl. Part. Phys. {\bf 34}, 2545 (2007).
\bibitem{Casten3} R. F. Casten, Phys. Rev. Lett. {\bf 54}, 1991
(1985).
\bibitem{Casten4} R. F. Casten, D. S. Brenner, and P. E. Haustein,
Phys. Rev. Lett. {\bf 58}, 658 (1987).
\bibitem{Cakirli} R. B. Cakirli and R. F. Casten, Phys. Rev. Lett.
{\bf 96}, 132501 (2006).
\bibitem{Casten5} R. F. Casten and N. V. Zamfir, Phys. Rev. Lett.
{\bf 70}, 402 (1993).
\bibitem{Zhao} Y. M. Zhao, R. F. Casten, and A. Arima, Phys. Rev.
Lett. {\bf 85}, 720 (2000).
\bibitem{Casten2} R. F. Casten and N. V. Zamfir, J. Phys. G {\bf
22}, 1521 (1996).
\bibitem{Heyde} K. Heyde, P. Vanisacker, R. F. Casten, and J. L. Wood, Phys. Lett. {\bf 155B}, 303 (1985).
\bibitem{Federmann} P. Federmann and S. Pittel, Phys. Lett. {\bf 69B}, 385 (1977).
\bibitem{Dobaczewski} J. Dobaczewski, W. Nazarewicz, J. Skalski, and T. Werner, Phys. Review Lett. {\bf 60}, 2254, (1988).
\bibitem{Stoitsov} M. Stoitsov, R. B. Cakirli, R. F. Casten, W. Nazarewicz, and W. Satula, Phys. Rev. Lett. {\bf 98}, 132502 (2007).
\bibitem{Ha1} E. Ha and D. Cha, J. Korean Phys. Soc. {\bf 50}, 1172 (2007).
\bibitem{Sabbey} B. Sabbey, M. Bender, G. F. Bertsch, and P.-H. Heenen, Phys. Rev. {\bf C}75, 044305 (2007).
\bibitem{Ha2} E. Ha and D. Cha, Phys. Rev. {\bf C}75, 057304 (2007).
\bibitem{Raman} S. Raman, C. W. Nestor, Jr., and P. Tikkanen, At.
Data Nucl. Data Tables {\bf 78}, 1 (2001).

\end{references}
\end{document}